\documentclass[aip,reprint]{revtex4-1}

\usepackage{graphicx}
\usepackage{array}
\usepackage{bm}
\usepackage[english]{babel}
\usepackage{dcolumn}
\usepackage{amssymb}
\usepackage{amsmath}

\begin{document}

\title{Bounds for Hamiltonians with arbitrary kinetic parts}

\author{Claude \surname{Semay}}
\email[E-mail: ]{claude.semay@umons.ac.be}
\affiliation{Service de Physique Nucl\'{e}aire et Subnucl\'{e}aire,
Universit\'{e} de Mons,
Acad\'{e}mie universitaire Wallonie-Bruxelles,
Place du Parc 20, 7000 Mons, Belgium}

\date{\today}

\begin{abstract}
A method is presented to compute approximate solutions for eigenequations in quantum mechanics with an arbitrary kinetic part. In some cases, the approximate eigenvalues can be analytically determined and they can be lower or upper bounds. A semiclassical interpretation of the generic formula obtained for the eigenvalues supports a new definition of the effective particle mass used in solid state physics. An analytical toy model with a Gaussian dependence in the momentum is studied in order to check the validity of the method. 
\end{abstract}


\maketitle

\section{Introduction}
\label{intro}

The envelope theory \cite{hall89,hall01,hall01b,hall02,hall03,hall05} is a powerful method to obtain approximate solutions, eigenvalues and eigenstates, of eigenequations in quantum mechanics. It has been rediscovered and developed recently under the name of the auxiliary field method \cite{silv08,silv09,eigen,silv11,sema11a,sema12}. Both techniques are equivalent \cite{buis09}, but they were introduced from completely different starting points. Let us assume that the Hamiltonian $H$ can be written as (in the following, we will work in natural units $\hbar = c = 1$)
\begin{equation}
\label{TpV}
H = T(p) + V( r),
\end{equation}
with $p=|\bm p|$ and $r=|\bm r|$. Such a form is relevant for one-body and two-body systems. The basic idea is to replace this Hamiltonian $H$ by another one $\widetilde H$ which is solvable, the eigenvalues of $\widetilde H$ being optimized to be as close as possible to those of $H$. Following the structure of the Hamiltonian, the approximate eigenvalues \emph{i)} can be upper or lower bounds, or not to have a variational character; \emph{ii)} can be obtained on a closed-form expression, or only numerically computed. In the most favorable case, an analytical bound, the dependence of the eigenvalues on parameters of the Hamiltonian and the quantum numbers can be determined, giving deep insights about the structure of the solutions and a reliable estimation of the spectrum. Even in the less favorable situation, a non-variational numerical approximation, it is possible to check easily and rapidly more elaborate numerical computations. 

At the origin, the method has been developed for Schr\"odinger equations \cite{hall89,silv08}, and afterwards
it has been extended for the semirelativistic kinematics \cite{hall01,silv09}. The purpose of this work is to generalize the technique to arbitrary kinetic operators. This is motivated by the existence of non-standard kinetic energies in some physical problems, for instance in atomic physics (non-parabolic dispersion relation) \cite{arie92} or in hadronic physics (particle mass depending on the relative momentum) \cite{szcz96}. Another motivation is to support the new definition of the effective particle mass proposed in Refs.~\cite{arie92,arie12,arie12b,arie12c}. 

The generic method to compute approximate solutions (which could be upper or lower bounds) of a Hamiltonian with a non-standard kinetic part is presented in Section~\ref{ge}. As this work is a generalization of the one described in Ref.~\cite{sema12}, the main equations are worked out without too many details. In Section~\ref{sc}, a semiclassical interpretation of the generic equations is given. In order to check the validity of the method, an analytical toy model with a Gaussian form for the kinetic part is solved and the formula obtained is compared with numerical solutions in Section~\ref{tm}. Concluding remarks are given in Section~\ref{conclu}.

\section{General equation}
\label{ge}

The envelope theory is generalized here to treat on the same footing the potential and kinetic parts. The idea is to replace the Hamiltonian~(\ref{TpV}) by the following one
\begin{equation}
\label{Htilde}
\widetilde H = \widetilde T + \widetilde V,
\end{equation}
with
\begin{equation}
\label{Ttilde}
\widetilde T = \frac{\bm p^2}{2\, \nu} + T(J(\nu)) - \frac{J(\nu)^2}{2\, \nu},
\end{equation}
and
\begin{equation}
\label{Vtilde}
\widetilde V = \rho\,P(r) + V(I(\rho)) -\rho\,P(I(\rho)).
\end{equation}
$\nu$ and $\rho$ are two real parameters, and we assume that the following functions are well-defined 
\begin{eqnarray}
\label{Htilde2}
&&\quad I(x)=K^{-1}(x), \quad K(x)=\frac{V'(x)}{P'(x)}, \nonumber \\
&&\quad J(x)=L^{-1}(x), \quad L(x)=\frac{x}{T'(x)}.
\end{eqnarray}
The kinematics of $\widetilde H$ is nonrelativistic, $P(x)$ is an auxiliary potential, and a prime denotes the derivative. An eigenvalue of this Hamiltonian is given by 
\begin{equation}
\label{EAFM}
E(\nu_,\rho) = T(J(\nu)) - \frac{J(\nu)^2}{2\, \nu} 
 +V(I(\rho)) - \rho\,P(I(\rho)) + \epsilon(\nu,\rho),
\end{equation}
where $\epsilon(\nu,\rho)$ is an eigenvalue of the nonrelativistic Hamiltonian
\begin{equation}
\label{HNR}
H_{\textrm{\scriptsize{NR}}} =  \frac{\bm p^2}{2 \nu}+ \rho\,P(r).
\end{equation}
Kinetic and potential parts are treated in a similar way, but with $x^2$ which is the counterpart of $P(x)$ and $1/(2\, \nu)$ which is the counterpart of $\rho$. The approximation for an eigenvalue of the genuine Hamiltonian is given by $E(\nu_0,\rho_0)$ for which 
\begin{equation}
\label{condextrema}
\left. \frac{\partial E(\nu,\rho)}{\partial \nu}\right|_{\nu_0,\rho_0} = \left. \frac{\partial E(\nu,\rho)}{\partial \rho}\right|_{\nu_0,\rho_0} = 0.
\end{equation}
Within these conditions, $\widetilde T$ is tangent to $T$ and $\widetilde V$ is tangent to $V$. The comparison theorem \cite{reed78,sema11b} implies that, if $\widetilde T \ge T$ and $\widetilde V \ge V$ for all values of the arguments, then the eigenvalues of $\widetilde H$ are upper bounds of the corresponding eigenvalues of $H$. Reciprocally, if $\widetilde T \le T$ and $\widetilde V \le V$ for all values of the arguments, then the eigenvalues of $\widetilde H$ are lower bounds. In other cases, no guarantee exists about the variational character of the approximations. A simple criteria to determine if a bound exists is given in Refs.~\cite{hall01,hall01b,hall02} for the potential, but it can be also used for the kinetic part. Let us define two functions $h$ and $g$ such that 
\begin{equation}
\label{hg}
T(x) = h(x^2) \quad \textrm{and}\quad V(x) = g(P(x)).
\end{equation}
If $h''(x)$ and $g''(x)$ are both concave (convex) functions, $E(\nu_0,\rho_0)$ is an upper (lower) bound of the genuine eigenvalue. If $T(p) \propto p^2$ ($V(r) \propto P(r)$), the variational character is solely ruled by the convexity of $g(x)$ ($h(x)$). 

Interesting results can be obtained if the auxiliary potential is a power law,
\begin{equation}
\label{Pr}
P(r) = \textrm{sgn}(\lambda)\, r^\lambda \quad \textrm{with} \quad 0 \ne \lambda > -2.
\end{equation}
Within this condition, $\rho$ and $\nu$ are always positive quantities, and $\epsilon(\nu,\rho)$ can be written under the form  \cite{silv08,hall89}
\begin{equation}
\label{epsPr}
\epsilon(\nu,\rho) = \frac{\lambda+2}{2 \lambda}\left( |\lambda|\rho \right)^{2/(\lambda+2)}\left(\frac{Q^2}{\nu}\right)^{\lambda/(\lambda+2)}, 
\end{equation}
where $Q$ is a global quantum number. The method is particularly interesting if $Q$ is exactly known. This is the case for the Coulomb interaction ($\lambda=-1$, $Q=n+l+1$) and the harmonic potential ($\lambda=2$, $Q=2\, n+l+3/2$). If $\lambda=1$, $Q$ is known only for $l=0$ states and is equal to $2 (-\alpha_n/3)^{2/3}$, where $\alpha_n$ is the $(n + 1)$th zero of the Airy function Ai. For arbitrary values of $\lambda$, simple and good analytical approximations can be found in Ref.~\cite{silv08}. But, if $Q$ is not computed with a sufficient accuracy, the variational character of a bound cannot be guaranteed. 

After some algebra, constraints (\ref{condextrema}), with $P(r)$ given by (\ref{Pr}), reduce to 
\begin{eqnarray}
\label{cond1}
(|\lambda|\, Q^\lambda)^\frac{2}{\lambda+2} (\rho_0\,\nu_0)^\frac{2}{\lambda+2} &=& p_0^2, \\
\label{cond2}
(|\lambda|\, Q^\lambda)^\frac{2}{\lambda+2} (\rho_0\,\nu_0)^{-\frac{\lambda}{\lambda+2}} &=& |\lambda|\, r_0^\lambda,
\end{eqnarray}
where $r_0=I(\rho_0)$ and $p_0=J(\nu_0)$. From (\ref{cond1}) and (\ref{cond2}), we can deduce that $r_0\, p_0 = Q$. Taking into account this result, plus the relations $\rho_0=K(r_0)=V'(r_0)/(|\lambda|\,r_0^{\lambda-1})$ and $\nu_0=L(p_0)=p_0/T'(p_0)$, (\ref{cond1}) and (\ref{cond2}) can be written in the more compact form (\ref{AFM3}).
After some algebra, the approximate eigenvalue $E(\nu_0,\rho_0)$ given by (\ref{EAFM}) with the parameterization (\ref{epsPr}) can be greatly simplified into (\ref{AFM1}). Finally, the approximate solution is given by the following set of equations
\begin{eqnarray}
\label{AFM1}
&&E = T(p_0)+V(r_0), \\
\label{AFM2}
&&p_0 = \frac{Q}{r_0}, \\
\label{AFM3}
&&p_0\, T'(p_0) = r_0\, V'(r_0).
\end{eqnarray}
The parameter $r_0$ can then be interpreted as a mean distance between the particles and $p_0$ as a mean momentum per particle. Both parameters depend on the quantum numbers via $Q$. The value $E$ can be a (upper or lower) bound following the convexity of functions $h(x)$ and $g(x)$, as explained above. Let us note that the only trace of the auxiliary potential is contained in the value of $Q$, and that (\ref{AFM3}) is the translation into the variables $r_0$ and $p_0$ of the generalized virial theorem \cite{virial}. The system~(\ref{AFM1})-(\ref{AFM3}) is similar to the systems (3.2)-(3.4) in Ref.~\cite{sema11a} and (15)-(17) in Ref.~\cite{sema12}, but here the form of the kinetic part is arbitrary. 

Let us note $|\nu_0,\rho_0\rangle$ an eigenstate of the nonrelativistic Hamiltonian $H_{\textrm{\scriptsize{NR}}}$ given by (\ref{HNR}) with $\nu=\nu_0=p_0/T'(p_0)$ and $\rho=\rho_0=V'(r_0)/(|\lambda|\,r_0^{\lambda-1})$. Such a state is an approximation of the corresponding eigenstate of $H$ \cite{eigen}.  Using the Hellmann-Feynman theorem \cite{Hell35} as in Ref.~\cite{silv08}, it can be shown that
\begin{eqnarray}
\label{pmean}
\langle \nu_0,\rho_0 | \bm p^2 | \nu_0,\rho_0\rangle &=& p_0^2,\\
\label{rmean}
\langle \nu_0,\rho_0 | r^\lambda |\nu_0,\rho_0\rangle &=& r_0^\lambda.
\end{eqnarray}
This confirms the interpretation of parameters $r_0$ and $p_0$ as mean values.

\section{semiclassical interpretation}
\label{sc}

\begin{figure}[ht]
\begin{center}
\includegraphics*[width=6cm]{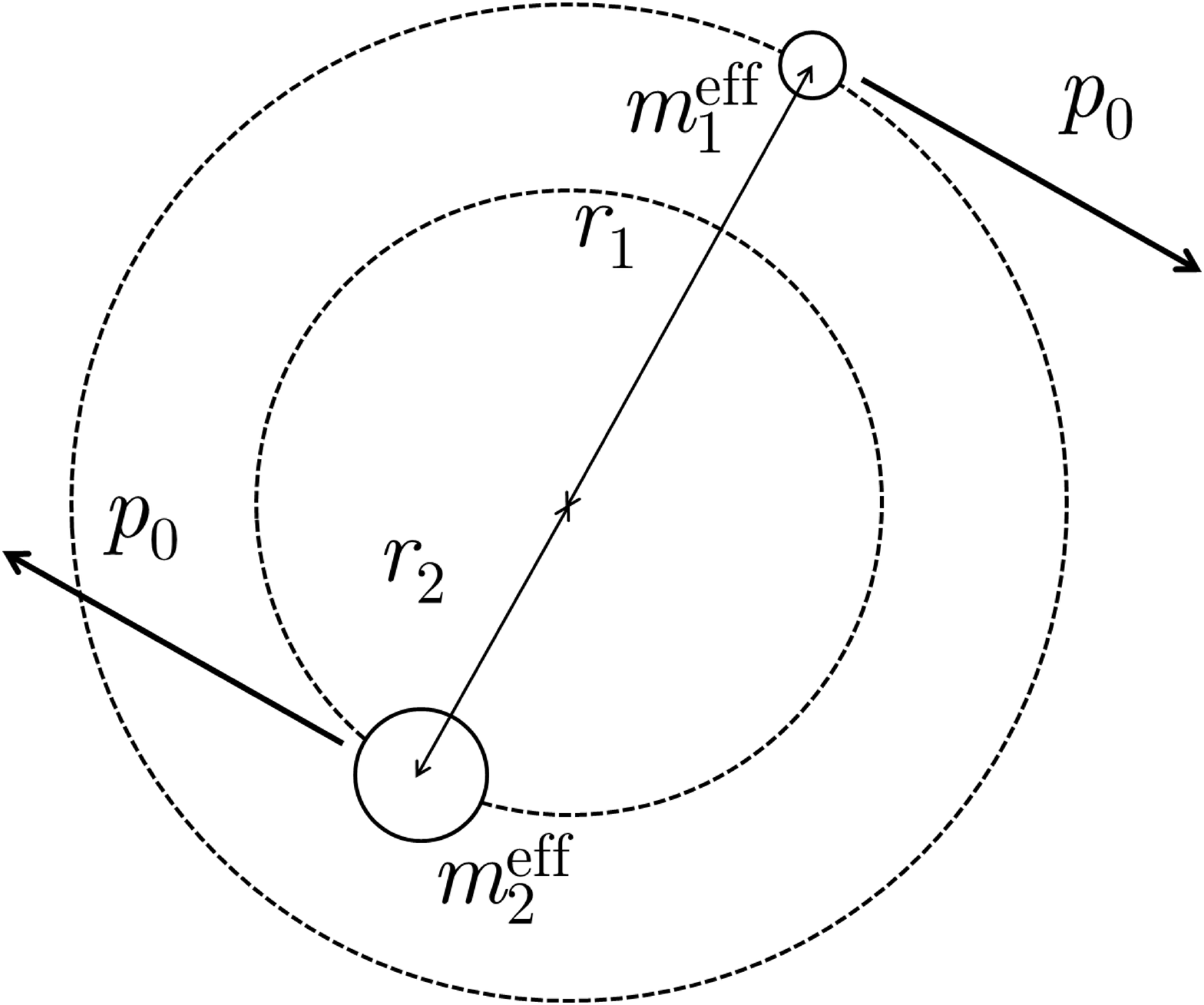}
\caption{Classical circular motion of two particles in their center of mass frame. \label{fig:class}}
\end{center}
\end{figure}

In Refs.~\cite{arie92,arie12,arie12b,arie12c}, an effective particle mass $m^\textrm{eff}$ is defined by the relation $\bm p = m^\textrm{eff}\, \bm v$, where $\bm v$ is the group velocity of the associated wave packet. It follows then that
\begin{equation}
\label{meffdef}
m^\textrm{eff}=p\,\left( \frac{d T}{d p} \right)^{-1}.
\end{equation}
Using this definition, a semiclassical interpretation of the system (\ref{AFM1})-(\ref{AFM3}) is possible. If $T_i(p_i)$ is the kinetic energy for the $i$th particle, the kinetic energy for two particles can be written $T_1(p_0)+T_2(p_0)=T(p_0)$, where $p_0$ is the module of the common momentum in the center of mass frame. With the effective mass (\ref{meffdef}), we have
\begin{equation}
\label{meff}
m_i^\textrm{eff}=\frac{p_0}{T_i'(p_0)},
\end{equation}
and the speed of the $i$th particle is given by $v_i=T_i'(p_0)$. As in Ref.~\cite{sema12}, let us assume a classical circular motion for the two particles (see Fig.~\ref{fig:class}). The centripetal force $F_i$ acting on the $i$th particle is given by
\begin{equation}
\label{Fi2}
F_i = m_i^\textrm{eff} \frac{v_i^2}{r_i} = p_0 \frac{T_i'(p_0)}{r_i}.
\end{equation}
If $r_0$ is the distance between the two particles, the rigid rotation constraints, $r_0=r_1+r_2$ and $v_1/r_1 = v_2/r_2$, imply that 
\begin{equation}
\label{r0r1r2}
r_0 = r_i \frac{T'(p_0)}{T_i'(p_0)}.
\end{equation}
If the force acting on the $i$th particle comes from the potential $V(r)$ generated by $j$, then $F_1=F_2=V'(r_0)$. (\ref{Fi2}) and (\ref{r0r1r2}) can be recast into the form (\ref{AFM3}), and it is obvious than (\ref{AFM1}) gives the mass of the system. A semiclassical quantification of the total orbital angular momentum gives $r_0\, p_0 = l+1/2$, and we obtain a system very similar to (\ref{AFM1})-(\ref{AFM3}). This supports (\ref{meffdef}) as a good definition for the effective mass.

\section{A toy model}
\label{tm}

In this section, we solve a simple toy model in order to check the relevance of the method. Let us consider the following Hamiltonian with a Gaussian form for the kinetic part
\begin{equation}
\label{TpVexp}
H = \sigma\, m\, \exp\left( \frac{\bm p^2}{2\, m^2}\right) + a\, r^2,
\end{equation}
where the parameter $m$ plays the role of a mass. Indeed, for high values of $m$ ($\gg a^{1/3}$), $H$ reduces to a harmonic oscillator Hamiltonian
\begin{equation}
\label{TpVexplim}
H \to \sigma\, m + \frac{\sigma\, \bm p^2}{2\, m} + a\, r^2 + \textrm{O}\left( \frac{\bm p^4}{m^2}\right).
\end{equation}
The parameter $\sigma=1$ or 2 is the number of particles (arbitrary positive value of $\sigma$ can also be considered to study duality relations between different many-body systems \cite{silv11}).
Such a Hamiltonian is not realistic but it has been chosen because \emph{i)} it admits an analytical lower bound (see below); \emph{ii)} it reduces to the well known case (\ref{TpVexplim}) in a well-defined limit;  \emph{iii)} accurate numerical solutions are not easy to obtain (see below). It is more convenient to work with the conjugate dimensionless variables $\bm x = \sqrt{a/(\sigma\, m)}\,\bm r$ and $\bm q = \sqrt{\sigma\, m/a}\,\bm p$ for the dimensionless Hamiltonian $H_{\textrm{d}}=H/(\sigma\, m)$ given by
\begin{equation}
\label{TpVexpred}
H_{\textrm{d}} = \exp\left( k\,\bm q^2 \right) + x^2 \quad \textrm{with} \quad k=\frac{a}{2\,\sigma\, m^3}.
\end{equation}
The corresponding eigenvalues are noted $\epsilon = E/(\sigma\, m)$. 

\begin{figure}[ht]
\begin{center}
\includegraphics*[width=0.45\textwidth]{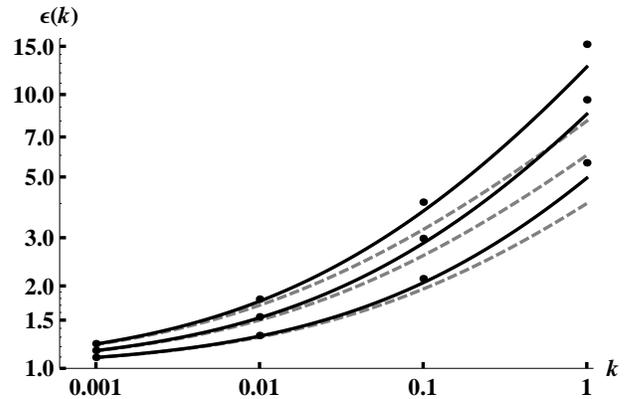}
\caption{Eigenvalues $\epsilon$ of (\ref{TpVexpred}) as a function of $k$. By increasing energy, the eigenvalues corresponds to $(n,l)=(0,0)$, $(0,1)$ and $(1,0)$. Dots: accurate numerical solutions; Solid black lines: lower bound (\ref{Eexpred}) with $Q=2n+l+3/2$; Dashed grey lines: harmonic oscillator approximation (\ref{Eexpredho}) with $Q=2n+l+3/2$. \label{fig:expp2}}
\end{center}
\end{figure}

Using the set of equations (\ref{AFM1})-(\ref{AFM3}), the following approximate solution can be found
\begin{equation}
\label{Eexpred}
\epsilon_\textrm{app} = \exp\left( 2\, W_0\left( \sqrt{k}\, \frac{Q}{2} \right) \right)\left[ 1+ 2\, W_0\left( \sqrt{k}\, \frac{Q}{2} \right) \right],
\end{equation}
where $W_0(z)$ is the main branch of the Lambert function (also called Omega function or product logarithm) \cite{corl96}. If $k \ll 1$, we can write $\epsilon_\textrm{app} = \epsilon_\textrm{HO} + \textrm{O}\left( k \right)$ with
\begin{equation}
\label{Eexpredho}
\epsilon_\textrm{HO} = 1+2\,\sqrt{k}\, Q.
\end{equation}
This corresponds to the harmonic oscillator approximation. A natural choice is to take $P(x)=V(x)=x^2$. Then, $Q=2n+l+3/2$ and the function $h(x)$ defined by (\ref{hg}) is convex. So, (\ref{Eexpred}) is a lower bound, whose quality is shown on Fig.~\ref{fig:expp2}. The numerical solutions of (\ref{TpVexpred}) have been computed with the the three-dimensional Fourier grid Hamiltonian method \cite{mars89,bali91,brau98} which is particularly well suited for this type of Hamiltonian. This numerical procedure is equivalent to an expansion in a special basis \cite{sema00} and implies the computation of Hamiltonian matrix elements. Because of the exponential function in (\ref{TpVexpred}), these matrix elements can be huge numbers and they must be computed with a very high accuracy to obtain stable and accurate eigenvalues. One can see that the lower bound is quite good and more accurate that the harmonic oscillator approximation when $k$ increases.

\section{Concluding remarks}
\label{conclu}

The envelope theory (or equivalently the auxiliary field method) is a method to compute approximate solutions (generally upper or lower bounds) of Hamiltonians of the form $H=T(p) + V( r)$, where the kinetic part $T$ is a nonrelativistic \cite{hall89,silv08} or a semirelativistic one \cite{hall01b,sema12}. In this paper, it is shown that the method can be used for arbitrary forms of $T$. The idea is to replace the Hamiltonian $H$ by another one $\widetilde H = \widetilde T + \widetilde V$ which is solvable, and with $\widetilde T$ and $\widetilde V$ respectively tangent to $T$ and $V$. Provided $\widetilde T$ is a nonrelativistic kinetic operator and $\widetilde V$ a power-law potential, the approximate eigensolutions can be easily computed by solving a set of three equations which have a natural semiclassical interpretation. Nevertheless, the computation is a full quantum one since eigenvalues and eigenstates are obtained for a well-defined global quantum number $Q$. 

A priori, This method can be applied to a wide variety of Hamiltonians relevant for various domains, from atomic to hadronic physics. With a good choice of the power law, the value of $Q$ is analytically known and the eigenvalues obtained can be upper or lower bounds of the genuine energies. If numerical approximations can be easily computed, closed-form formulae can even be obtained for some particular Hamiltonians. The toy model studied here with a harmonic potential and a Gaussian kinetic operator is in the most favorable situation: A quite accurate analytical lower bound is obtained giving deep insights about the structure of the solutions.

\begin{acknowledgments}
C.S. thanks the F.R.S.-FNRS for financial support. 
\end{acknowledgments}

\end{document}